\documentclass[prl,twocolumn,showpacs]{revtex4}

\usepackage{amsmath}
\usepackage{amssymb}

\usepackage{graphicx}

\begin{document}

\title{Resonance magneto-resistance in double barrier structure
with spin-valve}

\author{N. Ryzhanova$^{1}$,G. Reiss$^{2}$ , F. Kanjouri$^{1,3}$,
A. Vedyayev$^{1}$ }
\address{
$^2$Fakult\"{a}t f\"{u}r Physik, Universit\"{a}t Bielefeld, 33501
Bielefeld,
    Germany \\
$^1$Department of Physics, Moscow Lomonosov University, Moscow
119899,
    Russia \\
$^3$Department of Physics, Yazd University, Yazd, Iran}

\begin{abstract}
The conductance and tunnel magneto-resistance (TMR) of the double
barrier
  magnetic tunnel junction with spin-valve sandwich (F/P/F)
  inserted between two insulating barrier, are theoretically investigated.
  It is shown, that resonant tunnelling, due to the quantum well states of the
  electron confined between two barriers, sharply depends on the mutual
  orientation of the magnetizations of ferromagnetic layers F.
  The calculated optimistic value of TMR exceeds 2000\% .
\end{abstract}

\maketitle

During the last  years a lot of attention is paid to the
investigation of magnetic tunnel junctions (MTJs) as promising
candidates for application in magnetic random access memory
(MRAM), read heads etc\cite{Wolf[2001],Moodera[1999]}. One of the
problems, what has to be resolved for practical applications, is
to reach higher value of the tunnel magneto-resistance (T.M.R) and
low resistance value.
 Zhang et al \cite{Zhang[1997]} suggested to use double barrier tunnel junctions
 (DBTMJs), which exhibit resonant tunnelling due to the formation of
 resonant level in the metallic spacer placed between two barriers, to
 increase the value of T.M.R (their calculated value of T.M.R reaches
 90\% ).
  In \cite{Vedyayev[1998]} it was shown that ideal spin-valve (TMR up to almost
  100\%
  )
   may be constructed by using triple barrier structure. However until
    now in experiment with  DBTMJ the value of the observed \cite{Montaigne[1998)]} TMR
    doesn't exceeds its value for a single barrier, what may be due to
     the suppression of resonance by roughness of interfaces in these
     structures\cite{Chshiev[2002]}.

     Similar structure was used for constructing spin-valve
     transistor\cite{Dieny[1995]},where spin-valve element was
     inserted between two Schottky barriers. however the transport
     in this device is due to hot elektrons with energy above the
     height of Schottky barrier, so the quantum well state do not
     affect the electron's transport.

\begin{figure}[h]
\includegraphics*[width=0.5\textwidth]{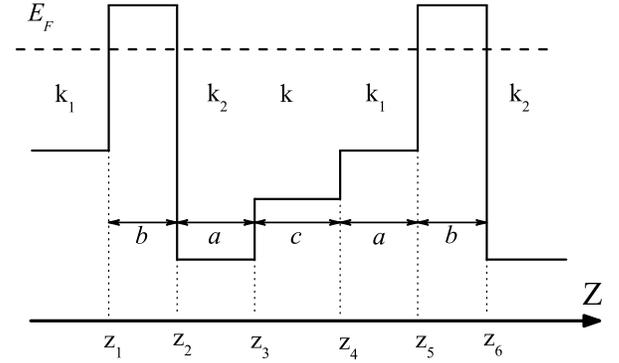}
\caption{The potential profile of DBTMJ with inserted spin-valve
for spin-up electron.} \label{fig:fig1}
\end{figure}

In the present paper we suggest to place between two insulating
barriers the metallic spin-valve sandwich (F1/P/F2), where F1, F2
are thin ferromagnetic metal layers and P-nonmagnetic metal
spacer. In this case the position of the resonant level in quantum
well between two barriers may be tuned by external magnetic field
what changes the mutual orientation of magnetizations in F1 and F2
layers. \\ The one band Hamiltonian of the multilayered structure
depicted on the (Fig. \ref{fig:fig1}) has in the ground states
form:
\begin{equation}
  \widehat{H}=\sum\limits_{\sigma,i}
  \frac{\widehat{P}^2}{2m}+U_{i}+\text{sign}\sigma\varepsilon_{i}\label{eq:Hamilton}
\end{equation}


 where $U_i$ is the bottom of the band in the
i-th layer and $\varepsilon_i$ is the exchange energy, different
from 0 in the ferromagnetic layers and equal 0 in nonmagnetic
ones, $\sigma=\pm 1$ is the spin index. To calculate the
conductance of the system we used the Kubo formula in mixed
$\kappa-z$ representation ($\kappa$  is the component of wave
vector in the plain of the layers and $z$ is the coordinate
perpendicular to this plain):
\begin{multline}
\sigma(z,z^\prime)=\frac{1}{\pi} \frac{e^2}{\hbar}
    \left(\frac{\hbar^2}{2m}\right)^2\nonumber
   \sum\limits_\kappa \left\{\left(\frac{\partial{G}_\kappa(z,z^\prime)}
   {\partial z}-\right. \right.\\ \left. \left.-
    \frac{\partial{G}_\kappa^{*}(z,z^\prime)}{\partial
    z}\right)
    \left(\frac{\partial{G}_\kappa(z,z^\prime)}{\partial z^\prime}-
    \frac{\partial{G}_\kappa^{*}(z,z^\prime)}{\partial
    z^\prime}\right)-\right. \\-
     \left.\left(\frac{\partial^2{G}_\kappa(z,z^\prime)}{\partial z\partial
    z^\prime}- \frac{\partial^2{G}_\kappa^{*}(z,z^\prime)}{\partial
    z\partial{z}^\prime}\right) \left({{G}_\kappa(z,z^\prime)}-
    {G}_\kappa^{*}(z,z^\prime)\right)\right\}\label{Kubo}
\end{multline}

 where $G(z,z^\prime)$   is the Green function of
the Hamiltonian (\ref{eq:Hamilton}) and it has to be found from
the solution of equation:
\begin{equation}
\left(E-\widehat{H}\right)G(z,z^\prime)=a_{0}\sigma(z,z^\prime)\label{eq:EQG}
\end{equation}

with  the condition of continuity of $G(z,z^\prime)$     and its
derivative on interfaces. The explicit expression for the Green
function has the form
\begin{multline}
  G(z,z^\prime)=2\left(\frac{ma_{o}}{\hbar^2}\right)
  \frac{e^{-ik_{1}(z-z_{1})}}{B(k_{1},k_{2})}\times\\\\
  \times\left(\Phi(k_{1},k_{2})
e^{-q(z^\prime-z_{2})}+\Psi(k_{1},k_{2})
e^{q(z^\prime-z_{2})}\right),\\\\
z<z_{1},\ \ \ z_{1}<z'<z_{2}.
\end{multline}
 where

\begin{eqnarray*}
\Phi(k_{1},k_{2})
 &=&e^{ik_{2}a}(q+ik_{2})\times\\&&\times\left\{(k+k_{2})\alpha
e^{ikc}-(k-k_{2})\beta e^{-ikc}\right\}+
\\&+&e^{-ik_{2}a}(q-ik_{2})\times\\&&\times\left\{(-k+k_{2})\alpha
e^{ikc}+(k+k_{2})\beta e^{-ikc}\right\},\\
\\\\\Psi(k_{1},k_{2})&=&e^{ik_{2}a}(q-ik_{2})\times\\&&\times\left\{(k+k_{2})\alpha
e^{ikc}-(k-k_{2})\beta e^{-ikc}\right\}+\\
&+&e^{-ik_{2}a}(q+ik_{2})\times\\&&\times\left\{(-k+k_{2})\alpha
e^{ikc}+(k+k_{2})\beta e^{-ikc}\right\},\\
\\B(k_{1},k_{2})&=&\left(q+ik_{1}\right)\Psi(k_{1},k_{2}) e^{-qb}-\\&& \ \ \
 \ \ \ \ \ -\left(q-ik_{1}\right)\Phi(k_{1},k_{2}) e^{qb}.
\end{eqnarray*}

\begin{eqnarray*}
\alpha&=&e^{ik_{1}a}(-\mu)(k+k_{1})+
e^{-ik_{1}a}(-\lambda)(k-k_{1}),\\
\beta&=&e^{ik_{1}a}(-\mu)(k-k_{1})+
e^{-ik_{1}a}(-\lambda)(k+k_{1}),\\
\mu&=&e^{-qb}(q+ik_{2})(q-ik_{1})-
e^{qb}(q-ik_{2})(q+ik_{1}),\\
\lambda&=&-e^{-qb}(q+ik_{2})(q+ik_{1})+
e^{qb}(q-ik_{2})(q-ik_{1}),
\end{eqnarray*}
$a_{0}$ lattice parameter, $q=\sqrt{q_{0}^2+\kappa^2}$, a
$k_{1,2}=\sqrt{k_{1,2F}^2-\kappa^2}$; $\kappa$ is the modulus of
electron momentum in the plain of the layers, $k_{1,2F}$ is the
Fermi vector for $1$ and $2$ spin electron, $q_{0}^{-1}$ is the
attenuation length inside the barrier.

The final expression for conductance is:
\begin{multline}
\sigma(k_{1},k_{2})=(\frac{64}{\pi}\sigma_{0})\\\int\kappa d\kappa
q Re(k_{1})
Im\left[\frac{-\Phi(k_{1},k_{2})\Psi^*(k_{1},k_{2})}{|B(k_{1},k_{2})|^2}\right]
\end{multline}
where $\sigma_{0}=\frac{2e^2}{h}=\frac{1}{12.9k\Omega}$.

\begin{figure}[h]
\includegraphics*[width=0.5\textwidth]{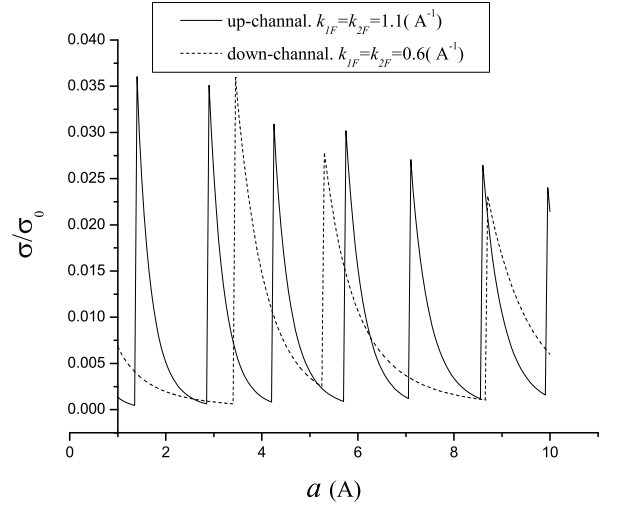}
\caption{The up and down spin channels conductances for parallel
alignment of magnetizations as a function of the middle
ferromagnetic layers thickness in the units $\frac{\sigma_{0}}{\mu
m^2}$ . $k=1$(\AA$^{-1}$),$ q_{0}=0.7$(\AA$^{-1}$), $b=15$(\AA),$
c=20$(\AA). } \label{fig:fig2}
\end{figure}


\begin{figure}[h]
\includegraphics*[width=0.5\textwidth]{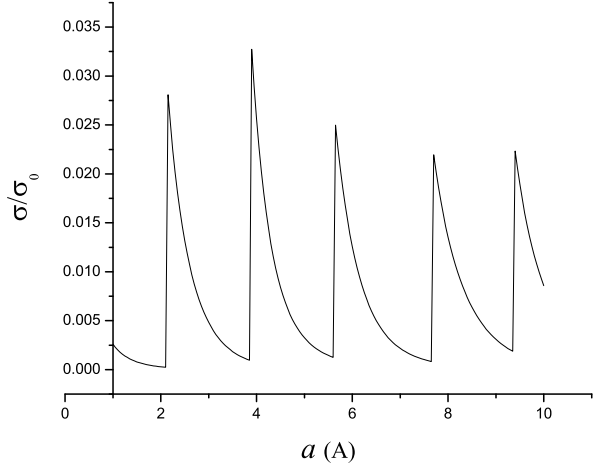}
\caption{The up and down spin channels conductances for
antiparallel alignment of magnetizations as a function of the
middle ferromagnetic layers thickness in the units
$\frac{\sigma_{0}}{\mu m^2}$.
$k_{1F}=1.1$(\AA$^{-1}$),$k_{2F}=0.6$(\AA$^{-1}$) For other
parameters see Fig. \ref{fig:fig2}. The relative orientation of
outer magnetisations will not affect the value of conductance too
much as it dose't change the position of resonance levels}
\label{fig:fig3}
\end{figure}

 The most important feature of the obtained expression for the conductance
is the pronounced resonances at point, where $Re (B) = 0$. These
resonances occurs for parallel and anti-parallel orientations of
magnetizations of F- layers at different thicknesses of F-layers,
so if resonance   takes place for example for parallel alignment,
it is no resonance for anti-parallel one and so the TMR may have
extremely high value. As it is clear from the (Fig.
\ref{fig:fig2},\ref{fig:fig3}) conductance as a function of the
middle F layers thicknesses for both channels (spin up and down)
has sharp maximums and the positions of these maximums are
different for parallel and anti-parallel orientation of the
magnetizations of F layers.


\begin{figure}[h]
\includegraphics*[width=0.5\textwidth]{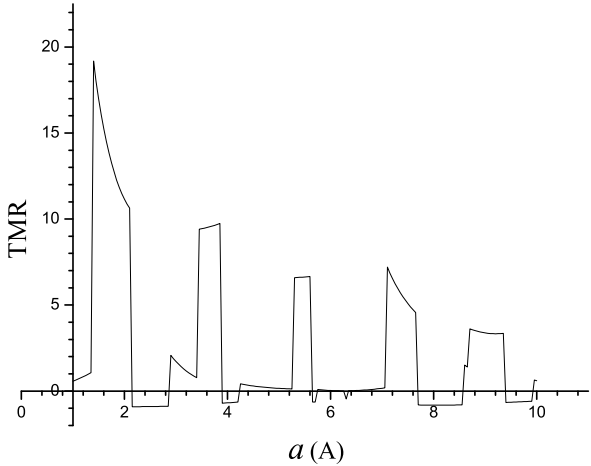}
\caption{ TMR as a function of the middle layer thickness.
$b=15$(\AA),$ c=20$(\AA).} \label{fig:fig4}
\end{figure}

 As a result the value of TMR (Fig. \ref{fig:fig4}) reaches extremely
high values. However these maximums
 may be smeared due to the fluctuation of the
thickness of the middle layer. We may notice that scattering of
electrons in the F layers depends on the electron's spin and this
scattering may give additional contribution  to the
magnetoresistance. However for thin ferromagnetic layers the
resistance of metallic layers   is negligible comparing to the
tunnel resistance and we may neglect this additional contribution
to the magnetoresistance.  So to observe the predicted resonances
in the case under consideration it is necessary to produce
structure with flat enough interfaces.

As it was shown in\cite{Chshiev[2002]} if the thickness of barrier
is fluctuating with deviation equal to $3a_{0}$ and the  area of
this fluctuation constitutes $5\%$ of the total area of the
junction the resonance in conductance are completely smeared.

This work was supported by the Russian fund for basic research.

\end{document}